\documentclass[aps,reprint,superscriptaddress,nofootinbib]{revtex4-1}

\usepackage{bm, amsfonts, amsmath, amssymb, xspace, mathtools}
\usepackage{graphicx}
\usepackage{hyperref}

\newcommand\rts{\ensuremath{\text{Red}\!\rightarrow\!\text{Syn}}\xspace}

\newcommand\korder{\ensuremath{k}\textsuperscript{th}-order\xspace}

\newcommand\phiid{\ensuremath{\Phi}ID\xspace}

\begin{document}

\title{Greater than the parts: A review of\\
the information decomposition approach to causal emergence}

\author{Pedro~A.M. Mediano}
\thanks{P.M. and F.R. contributed equally to this work.\\E-mail: pam83@cam.ac.uk, f.rosas@imperial.ac.uk}
\affiliation{Department of Psychology, University of Cambridge}
\affiliation{Department of Psychology, Queen Mary University of London}

\author{Fernando~E. Rosas}
\thanks{P.M. and F.R. contributed equally to this work.\\E-mail: pam83@cam.ac.uk, f.rosas@imperial.ac.uk}
\affiliation{Centre for Psychedelic Research, Imperial College London}
\affiliation{Data Science Institute, Imperial College London}
\affiliation{Centre for Complexity Science, Imperial College London}

\author{Andrea~I. Luppi}
\affiliation{University Division of Anaesthesia, University of Cambridge}
\affiliation{Department of Clinical Neurosciences, University of Cambridge}
\affiliation{Leverhulme Centre for the Future of Intelligence, University of Cambridge}

\author{Henrik~J. Jensen}
\affiliation{Centre for Complexity Science, Imperial College London}
\affiliation{Department of Mathematics, Imperial College London}
\affiliation{Institute of Innovative Research, Tokyo Institute of Technology}

\author{Anil~K. Seth}
\affiliation{Sackler Centre for Consciousness Science, University of Sussex}
\affiliation{CIFAR Program on Brain, Mind, and Consciousness, Toronto}

\author{Adam~B. Barrett}
\affiliation{Sackler Centre for Consciousness Science, University of Sussex}
\affiliation{The Data Intensive Science Centre, University of Sussex}

\author{Robin~L. Carhart-Harris}
\affiliation{Centre for Psychedelic Research, Imperial College London}
\affiliation{Psychedelics Division, Neuroscape, Department of Neurology, University of California San Francisco}

\author{Daniel Bor}
\affiliation{Department of Psychology, University of Cambridge}
\affiliation{Department of Psychology, Queen Mary University of London}

\begin{abstract}

Emergence is a profound subject that straddles many scientific disciplines,
including the formation of galaxies and how consciousness arises from the
collective activity of neurons. Despite the broad interest that exists on this
concept, the study of emergence has suffered from a lack of formalisms that
could be used to guide discussions and advance theories. Here we summarise,
elaborate on, and extend a recent formal theory of causal emergence based on
information decomposition, which is quantifiable and amenable to empirical
testing. This theory relates emergence with information about a system's
temporal evolution that cannot be obtained from the parts of the system
separately. This article provides an accessible but rigorous introduction to
the framework, discussing the merits of the approach in various scenarios of
interest. We also discuss several interpretation issues and potential
misunderstandings, while highlighting the distinctive benefits of this
formalism.

\end{abstract}

\maketitle

\section{Introduction}

Emergence is a key concept in several challenging open questions in science and
philosophy, and a subject of long-standing debate. A distinctively
controversial topic, research on emergence has been characterised by differing
assumptions and positions --- explicit and implicit --- about its nature and
role within science. At one extreme of the spectrum, \emph{reductionism} claims
that all that is `real' can always be explained based on sufficient knowledge
of a system's smallest constituents, and that coarse-grained explanations are
mere byproducts of our limited knowledge and/or computational ability. At the
other extreme, strong forms of \emph{emergentism} argue for a radical
independence between layers of reality, such that some high-level phenomena are
in principle irreducible to their low-level constituents.

Modern scientific practice is dominated by reductionist assumptions, at least
in its overall theoretical and philosophical commitments. At the same time, the
hierarchical organisation and in-practice relative independence of the domains
of different scientific disciplines (e.g. physics, biology) suggests that some
form of emergentism remains in play. There is therefore a need to formulate
principled, rigorous, and consistent formalisms of emergence, a need that is
especially pressing for those topics where strong emergentism retains intuitive
appeal --- such as the relationship between consciousness and the brain.

Riding on a wave of renewed philosophical
investigations~\cite{cunningham2001reemergence,bedau2008emergence}, recent work
is opening a new space of discussion about emergence that is firmly within the
realm of empirical scientific
investigation~\cite{graben2009stability,seth2010measuring,hoel2013quantifying,klein2020emergence,rosas2020reconciling,varley2021emergence,barnett2021dynamical}.
This work is developing formal principles and analytical models, which promise
to facilitate discussions among the community of interested researchers.
Moreover, having a formal theory of emergence will allow scientists to
formulate rigorous, falsifiable conjectures about emergence in different
scenarios and test them on data.

This article presents an overview of a recently proposed formal theory of
causal emergence~\cite{rosas2020reconciling} based on the framework of Partial
Information Decomposition (PID)~\cite{williams2010nonnegative}. In contrast
with other proposals, this approach is primarily \emph{mereological}: emergence
is considered to be a property of part-whole relationships within a system,
which depends on the relationship between the dynamics of parts of the system
and macroscopic features of interest. In what follows, we outline the necessary
mathematical background, present the core principles of the theory, and review
some of its key properties and applications.

\section{Technical preliminaries}
\label{sec:preliminaries}

\subsection{An information-centric perspective on complex systems}

Information theory is deeply rooted in probability theory, to the extent that
the axiomatic bases of both are formally equivalent~\cite{jizba2020shannon}.
Both approaches, in turn, are illuminated by the seminal work of E.T. Jaynes on
the foundations of thermodynamics~\cite{jaynes2003probability}, which proposes
that probability theory can be understood as an extension of Aristotelian logic
that applies to scenarios of partial or incomplete knowledge. In this context,
probability distributions are to be understood as epistemic statements used to
represent states of limited knowledge, and Shannon's entropy corresponds to a
fundamental measure of uncertainty.

This perspective leads to principled and broadly applicable interpretations of
information-theoretic quantities. In fact, while information theory was created
to solve engineering problems in data
transmission~\cite{shannon1948mathematical}, modern approaches cast information
quantities as measures of belief-updating in statistical
inference~\cite{Ince2017,barnett2012transfer,cliff2016information}. In this
view, measuring the mutual information between two parts of a complex system
does not require assuming one is `sending bits' to the other over some channel
--- instead, mutual information can be seen as the strength of the evidence
supporting a statistical model in which the two parts are coupled. Furthermore,
information-theoretic tools are widely applicable in practice, spanning
categorical, discrete and continuous, as well as linear and non-linear
scenarios. In fact, a wide variety of estimators and open-source software is
available, whose diversity in terms of assumptions and requirements allows
flexible and reliable calculations on a broad range of practical
scenarios~\cite{lizier2014jidt,james2018dit,novelli2019large}.

Together, these properties place information theory as a particularly
well-suited framework to study interdependencies in complex systems,
establishing information as a `common currency' of interdependence that allows
one to assess and compare diverse systems in a principled and
substrate-independent
manner~\cite{crutchfield2003regularities,lizier2012local,rosas2016understanding}.

\subsection{The fine art of information decomposition}\label{sec:PID}

Shannon's information is particularly useful for the study of complex systems
due to its decomposability. For example, the information about a variable $Y$
provided by two predictors $X_1$ and $X_2$, denoted by $I(X_1,X_2;Y)$, can be
decomposed via the \emph{information chain-rule}~\cite{cover1999elements} as
\begin{equation}\label{eq:chain_rule}
    I(X_1,X_2;Y) = I(X_1;Y) + I(X_2;Y|X_1),
\end{equation}
where $I(X_1;Y)$ corresponds to the information provided by $X_1$, and
$I(X_2;Y|X_1)$ refers to the information provided by $X_2$ given $X_1$ is
already known. Taking this idea one step further, the \emph{Partial Information
Decomposition} (PID) framework~\cite{williams2010nonnegative} proposes to
decompose each of these terms into \emph{information atoms} as follows:
\begin{equation}\label{eq:PID}
\begin{aligned}
    I(X_1;Y) =& \; \text{Red}(X_1,X_2;Y) + \text{Un}(X_1;Y|X_2) ~ ,\\
    I(X_2;Y|X_1) =& \; \text{Un}(X_2;Y|X_1)+\text{Syn}(X_1,X_2;Y) ~ ,
\end{aligned}
\end{equation}
where $\text{Red}(X_1,X_2;Y)$ represents the \emph{redundant} information about
$Y$ that is contained in both $X_1$ and $X_2$, $\text{Un}(X_1;Y|X_2)$ and
$\text{Un}(X_2;Y|X_1)$ correspond to the \emph{unique} information that is
conveyed by $X_1$ or $X_2$ but not the other, and $\text{Syn}(X_1,X_2;Y)$
refers to the \emph{synergistic} information that is provided by $X_1$ and
$X_2$ together but not by each of them separately.

To illustrate, consider our two eyes as sources of visual information about the
environment. The information that we still have when we close either eye is
redundant (for instance, information about colour), while the extra information
we derive from combining them (e.g. stereoscopic information about depth) is
synergistic. For further reading on PID, we refer the reader to
Refs.~\cite{williams2010nonnegative,wibral2017partial,timme2018tutorial}.

\subsection{Decomposing information dynamics:\\From PID to \phiid}

As a final piece of mathematical background, we now show how information
decomposition can be applied to the temporal evolution of a dynamical system.
Let's consider two interdependent processes sampled at times $t$ and $t'>t$,
and denote their corresponding values as $X_t^1, X_t^2$ and $X_{t'}^1,
X_{t'}^2$, respectively. The information that these two processes carry
together from $t$ to $t'$ is given by the \emph{time-delayed mutual
information} ($\text{TDMI}$), denoted by $I(\bm X_{t};\bm X_{t'})$ where $\bm
X_{t} = (X_{t}^1, X_{t}^2)$. By regarding $X_t^1$ and $X_t^2$ as predictors and
the joint future state $\bm X_{t'}$ as target, Eqs.~\eqref{eq:chain_rule} and
\eqref{eq:PID} allow us to decompose the $\text{TDMI}$ as follows:
\begin{align*}
    \text{TDMI} =
    \text{Red}(X_t^1,X_t^2;\bm X_{t'}) 
    + \text{Syn}(X_t^1,X_t^2;\bm X_{t'}) \\
    + \text{Un}(X_t^1;\bm X_{t'}|X_t^2) 
    + \text{Un}(X_t^2;\bm X_{t'}|X_t^1) ~ .
\end{align*}
However, this decomposition considers the future state as a single entity and,
hence, cannot discriminate between the various ways in which the predictors
affect different parts of the target.

This important limitation is overcome by a finer decomposition, called
\emph{Integrated Information Decomposition} (\phiid)~\cite{mediano2019beyond},
which establishes information atoms not only in terms of the relationship
between the predictors, but also between the targets. For example, information
can be carried redundantly by $X^1_t,X^2_t$ but received synergistically by
$X^1_{t'},X^2_{t'}$, which corresponds to a \phiid atom denoted (in simplified
notation) by \rts.

By considering these dynamical information atoms, \phiid establishes a way of
decomposing PID atoms into a sum of finer \phiid atoms. In particular,
each of the four PID atoms can be decomposed into four \phiid atoms, which
brings a decomposition of the $\text{TDMI}$ into $4\times 4=16$ distinct atoms.
For more details about the interpretation of each of the \phiid atoms, and
their generalisation to more than two time series, we refer the reader to
Refs.~\cite{mediano2019beyond,luppi2021like}.

\section{Formalising mereological causal emergence}
\label{sec:theory}

\begin{figure*}[t]
  \centering
  \includegraphics[width=0.85\textwidth]{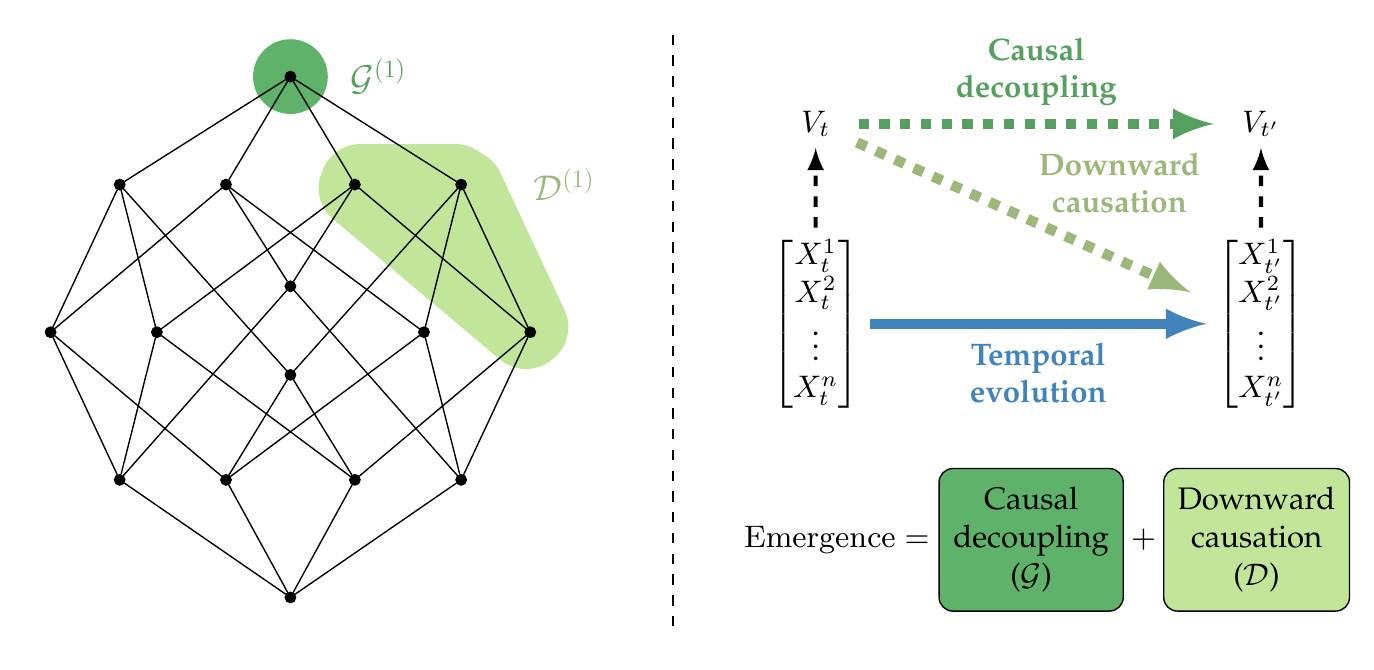}
  \caption{\textbf{Schematic of the \phiid approach to causal emergence}.
  (\emph{left}) Lattice of \phiid information atoms, with atoms corresponding
to causal decoupling ($\mathcal{G}$) and downward causation ($\mathcal{D}$)
highlighted. (\emph{right}) Relationship between system variables $\bm X_t$,
supervenient variables $V_t$, and emergent properties (c.f.
Eq.~\eqref{eq:modes_of_emergence}). Images adapted from
Refs.~\cite{mediano2019beyond,rosas2020reconciling,luppi2021like}.}
  \label{fig:diagram}
\end{figure*}

The first step towards using \phiid to formalise causal emergence is to
formalise the notion of \emph{supervenience} --- i.e. that one property `rests'
on another. For this purpose, one says that a variable $V_{t}$ is supervenient
on the state of the system $\bm X_{t}$ if it is a (possibly noisy) function of
$\bm X_{t}$. This definition implies that for $t<t'$ there is nothing about
$V_{t'}$ that can be predicted from the system's previous states $\bm X_{t}$
that cannot also be predicted from the system's current state $\bm X_{t'}$.

Building on this definition, a supervenient feature $V_t$ is said to exhibit
\emph{causal emergence of order $k$} if it has predictive power about the
future evolution of the underlying system $\bm X_t = (X_t^1, ..., X_t^n)$ that
is \korder unique with respect to the state of each part of the system, i.e. if
\begin{equation}\label{eq:unique_k}
    \text{Un}^{(k)}(V_t;\bm X_{t'}|\bm X_t) >0~.
\end{equation}

The notion of \korder unique information comes from a PID of $n$ predictors,
which generalises the case of two predictors discussed in the previous
section.\footnote{For scenarios with $n$ predictors, PID establishes the
following decomposition in terms of \korder redundancy, synergy, and unique
information~\cite[Lemma~2]{rosas2020reconciling}:
\unexpanded{\begin{align*}
    I(\bm X^n;Y) = 
    \; \text{Red}^{(k)}(\bm X^n;Y) +\text{Syn}^{(k)}(\bm X^n;Y) \\
    + \sum_{\beta\in\mathcal{B}_k} \text{Un}^{(k)}(\bm X^{\bm\beta};Y|\bm X^{-\bm\beta})~,
\end{align*}}
where $\bm X^n = (X_1,\dots,X_n)$ is a shorthand notation for the vector of all
predictors, $\bm X^{\bm \beta}$ is the vector of all predictors with indices
within the array $\bm\beta=(\beta_1,\dots,\beta_L)$, $\bm X^{-\bm \beta}$ is
the vector of all predictors with indices not in $\bm\beta$, and
$\mathcal{B}_k$ stands for the collection of subsets of $\{1,\dots,n\}$ of
cardinality $k$ or less.}
Intuitively, the \korder unique information $\text{Un}^{(k)}(V_t; \bm
X_{t'}|\bm X_t)$ is the information about $\bm X_{t'}$ that $V_t$ has access to
and no subset of $k$ or fewer parts of $\bm X_t$ has access to on its own
(although bigger groups may). Causal emergence is therefore defined as the
capability of some supervenient feature to provide predictive power that cannot
be reduced to underlying microscale phenomena --- up to order $k$. Put simply,
emergent features have more predictive power than their constituent parts.

Crucially, this framework accommodates the coexistence of supervenience and
irreducible predictive power of emergence, which have been previously thought
as paradoxical~\cite{bedau1997weak,bedau2002downward}. It does so by leveraging
the temporal dimension, such that supervenience is operationalised in terms of
\textit{instantaneous} relationships (between the system and its observables)
and emergence in terms of predictive power \textit{across time}. In this
context, a feature could be supervenient without being causally emergent, but
not \textit{vice versa}.

One of the main consequences of this theory is that, under relatively general
assumptions~\cite{rosas2020reconciling}, a system's capability to display
causally emergent features depends directly on how synergistic the system's
dynamics are. Specifically, a system $\bm X_t$ possesses causally emergent
features of order $k$ if and only if $\text{Syn}^{(k)}(\bm X_t; \bm
X_{t'})>0$~\cite[Theorem~1]{rosas2020reconciling}. Intuitively,
$\text{Syn}^{(k)}(\bm X_t; \bm X_{t'})$ is the information about the future
evolution that is provided by the whole system, but is not contained in any set
of $k$ or fewer predictors when considered separately from the rest.

This result has two important implications. First, the dependence of emergence
on synergistic dynamics suggests one can interpret the term
$\text{Syn}^{(k)}(\bm X_t; \bm X_{t'})>0$ as the \emph{emergence capacity} of a
system. Secondly, we can use the formal apparatus of \phiid to decompose
$\text{Syn}^{(k)}$ and distinguish two qualitatively different types of
emergence:
\begin{itemize}
    \item[(i)] \emph{Downward causation}, where an emergent feature has unique
    predictive power over specific parts of the system. Technically, a
    supervenient feature $V_t$ exhibits downward causation of order $k$ if
    $\text{Un}^{(k)}(V_t;\bm X^{\bm \alpha}_{t'}|\bm X_t) >0$ for some subset
    of $k$ time-series denoted by $\bm X_{t'}^{\bm \alpha}$.
    \item[(ii)] \emph{Causal decoupling}, in which an emergent feature $V_t$
    has unique predictive power not over any constituent of size $k$ or less,
    but on the system as a whole. Technically,  a supervenient feature $V_t$
    exhibits causal decoupling of order $k$ if $\text{Un}^{(k)}(V_t;V_{t'}|\bm
    X_t,\bm X_{t'}) >0$. This corresponds to `persistent synergies,' involving
    macroscopic variables that have causal influence on other macroscopic
    variables, above and beyond the microscale effects.
\end{itemize}
Further derivations show that a system has features that exhibit \korder
downward causation if and only if $\mathcal{D}^{(k)}(\bm X_t; \bm X_{t'})>0$,
and has \korder causally decoupled features if and only if
$\mathcal{G}^{(k)}(\bm X_t; \bm X_{t'}) >0$. Moreover, the \phiid framework
shows that this taxonomy of emergent phenomena is exhaustive, as emergence
capacity of a system can be decomposed as
\begin{equation}\label{eq:modes_of_emergence}
    \text{Syn}^{(k)}(\bm X_t; \bm X_{t'}) 
    = \mathcal{D}^{(k)}(\bm X_t; \bm X_{t'}) 
    + \mathcal{G}^{(k)}(\bm X_t; \bm X_{t'})~.
\end{equation}

A final aspect of this theory worth highlighting is that it provides practical
measures that are readily computable in large systems. In general, the value of
the terms in Eqs.~\eqref{eq:unique_k} and \eqref{eq:modes_of_emergence} depends
on a choice of redundancy function,\footnote{Multiple redundancy functions
exist, and ongoing work is exploring the strengths and weaknesses of different
choices. For more information, see Ref.~\cite{mediano2019beyond} and the
extensive PID literature.} whose estimation often requires large amounts of
data when the number of parts of the system grow. Fortunately, the \phiid
formalism of causal emergence allows to derive simple measures that provide
sufficient criteria for guaranteeing the presence of emergence and are
independent of the choice of redundancy function. Importantly, these measures
are relatively easy to calculate, as they avoid the `curse of dimensionality'
since they rely only on statistics related to low-order marginals. This key
feature allows the framework to be applicable on a wide range of scenarios, as
illustrated by the applications reviewed in Section~\ref{sec:applications}.
More information about these measures can be found in
Ref.~\cite{rosas2020reconciling}.

\section{Interpretation and remarks}
\label{sec:interpretation}

Having considered the main technical elements of the formalism, this section
discusses some key aspects of its interpretation while clarifying some
potential misunderstandings.

\subsection{Interventionist vs probabilistic causation}

Some interpretations (e.g. Ref.~\cite{dewhurst2021causal}) of the presented
framework place emphasis on its relation to Granger causality, as the
definition of causal emergence is based on predictive ability --- as opposed
to, for example, interventionist approaches to causality based on
counterfactuals, as proposed by Pearl~\cite{pearl2018book}. However, it is
important to note that the framework presented here belongs to neither the
Granger nor Pearl schools of thought, and admits both kinds of causal
interpretation depending on the underlying probability distribution from which
the relevant quantities are computed. As a matter of fact, all the quantities
described in Sec.~\ref{sec:theory} and Ref.~\cite{rosas2020reconciling} depend
only on the joint probability distribution $p(\bm X_{t'}, \bm X_t)$. If this
distribution is built using a conditional distribution $p(\bm X_{t'}|\bm X_t)$
that is equivalent to a \texttt{do()} distribution in Pearl's
sense~\cite{pearl2018book}, and the system satisfies a few other
properties,\footnote{Technically known as faithfulness and causal Markov
conditions --- see Ref.~\cite{koller2009probabilistic} for a detailed
description.} then the results of \phiid can be interpreted in an
interventionist causal sense. On the other hand, if the distribution is built
on purely observational data, then the decomposition obtained from \phiid
generally should be understood in the Granger-causal sense (i.e. as referring
to predictive ability). In both cases, the formalism developed here applies
directly, and it is only the interpretation of the findings that needs to be
adapted.

It is also important to clarify that the reason why correlation between
variables of a system of interest often does not imply causation is because of
hidden (i.e. unobserved) variables. However, if all the relevant variables are
measured, then Granger- and Pearl-type analyses coincide. Therefore, we
emphasise that while some results might not have an intervention-type
interpretation, this is not due to limitations of the formalism in principle
but only due to limitations of measurement in practice.

\subsection{Lack of invariance under change of coordinates}

A possible objection to the framework outlined here is that it critically
depends on the specific partition of the underlying system, i.e. in how the
\emph{parts} are defined. Put differently, synergy and unique information are
not invariant under changes in the way the micro-elements are construed ---
what is technically known as `change of coordinates.'\footnote{As a simple
example, consider the XOR gate $Y = X^1 \oplus X^2$, with $X^1,X^2$ i.i.d.
unbiased coin flips, and the change of coordinates $(Z^1,Z^2) = (X^1 \oplus
X^2, X^1)$. In this case $\text{Syn}(X^1, X^2; Y) = 1$, while $\text{Un}(Z^1; Y
| Z^2) = 1$, showing that information atoms are not invariant under changes of
coordinates in general.}

It is important to remark that this lack of invariance is not a bug, but rather
a feature of our framework. Recall that our theory is fundamentally a
\emph{mereological} one --- i.e. about the relationship between the whole and
its parts. Therefore, it is only natural that if the parts change,
quantification of the part-whole relationships observed in the system should
change too. Put differently, it is reasonable to expect that a mereological
account of emergence should critically depend on how the parts are defined, and
that any conclusions should be able to change if those parts change.

Following on from Sec.~\ref{sec:preliminaries}, we highlight that this property
aligns well with the epistemic interpretation of probabilities spearheaded by
Jaynes~\cite{jaynes2003probability}. If one embraces the idea that
probabilistic descriptions are representations of states of knowledge, then it
follows that the coordinates used to describe the system determine how the
joint distribution ought to be marginalised --- which is also part of our state
of knowledge. Then, it is to be expected that changing the system's coordinates
should change any conclusions drawn from the relationship between marginals ---
including causal emergence.

\subsection{On the order and scale of emergence}
\label{sec:order}

Although most of the empirical results from \phiid presented in the literature
so far (reviewed in the next section) correspond to emergence of order $k=1$,
it is important to highlight that the formalism allows us to tune the value of
$k$ to detect emergence at various spatial scales. In fact, being \korder
emergent implies that there is predictive ability related to interactions of
order $k+1$ or more. Therefore, while 1\textsuperscript{st}-order emergence
encompasses any pairwise information that could have predictive value, this can
be refined by searching for the relevance of higher orders. In this regard, it
is to be noted that a \korder emergent feature is emergent for all orders
$j<k$, and hence increasing the order makes finding emergent features
increasingly more challenging.

A related potential misunderstanding is to believe that the \phiid framework
for causal emergence only concerns predictive ability at the microscale,
without establishing a proper comparison with a
macroscale~\cite{varley2021emergence}. It is important to clarify that this
approach to emergence is established in terms of supervenient macroscopic
variables, which may be considered emergent depending on their dynamics and
predictive power over the evolution of the system --- not too dissimilar from
other approaches~\cite{hoel2013quantifying,varley2021emergence}. The fact that
dynamical synergy enables the existence of such emergent variables is not an
assumption, but a consequence of the theory. Moreover, this result enables a
powerful method to characterise emergence: unlike other theories, the \phiid
approach to causal emergence can determine the overall capability of a system
to host emergent properties without the need to specify any particular
macroscopic variable. Further, the `scale' of emergence is tuned by the
emergence order $k$, which sets the measures to focus on high-order
interdependencies that do not play a role at scales smaller than
$k+1$~\cite{mediano2019beyond}.

\section{Applications}\label{sec:applications}

Despite its recent inception, the presented framework has already proven
capable of providing insights about a wide range of phenomena. In the
following, we first present case studies that demonstrate how the framework
aligns with paradigmatic examples of putative emergent behaviour, and then
discuss recent results related to the human brain.

This framework provides two approaches to assess emergence in practice: one can
(i) test if a given feature of interest has emergent behaviour either directly
with the definition (Eq.~\ref{eq:unique_k}) or via the practical criteria
discussed at the end of Section~\ref{sec:theory}, or one can (ii) calculate the
capacity of a system to host \emph{any} emergent feature by computing its
dynamical synergy. The latter approach is more encompassing, but requires one
to use a redundancy function (see Section~\ref{sec:theory}) and usually scales
poorly with number of parts --- making its calculation in large systems very
challenging. The former approach focuses on a particular feature, but
circumvents those problems allowing one to deal with large systems. In the
following, the case studies reviewed in Section~\ref{sec:confirm} use the
practical criteria (i.e. not requiring a choice of redundancy function), while
most in Section~\ref{sec:fmri_studies} calculate dynamical synergy (i.e.
requiring a specific redundancy function).

\begin{figure*}[t!]
  \centering
  \includegraphics[width=0.85\textwidth]{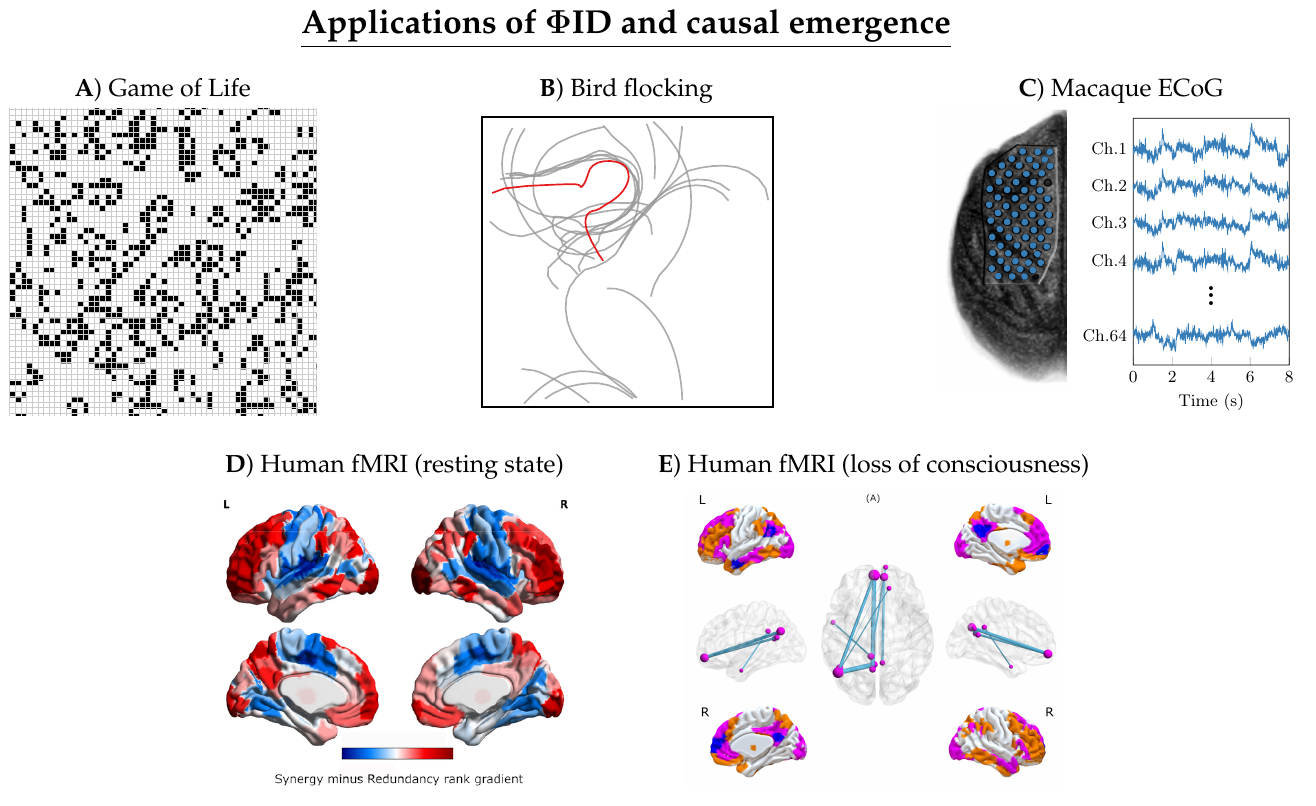}

  \caption{\textbf{Example published applications of the \phiid approach to
causal emergence}. Examples include \textbf{A)} Conway's Game of Life,
\textbf{B)} a bird flocking model, \textbf{C)} macaque ECoG during motor
control~\cite{rosas2020reconciling}, \textbf{D)} human resting-state fMRI brain
activity~\cite{luppi2020synergistic}, and \textbf{E)} human fMRI during loss of
consciousness~\cite{luppi2020synergisticb}. Images reproduced from
Refs.~\cite{rosas2020reconciling,luppi2020synergistic,luppi2020synergisticb}
and the \href{http://www.neurotycho.org}{Neurotycho} database.}

  \label{fig:applications}
\end{figure*}

\subsection{Confirming intuitions: Emergence in the\\Game of Life and bird flocks}\label{sec:confirm}

The efficacy of the presented framework to detect emergence was demonstrated in
a paradigmatic example of emergent behaviour: Conway's celebrated Game of Life
(GoL)~\cite{conway1970game}. In GoL, simple local rules determine whether a
given cell of a 2D grid will be ON (alive) or OFF (dead) based on the number of
ON cells in its immediate neighbourhood. The simple GoL rule nevertheless
results in highly complex behaviour, with recognisable self-sustaining
structures --- known as ``particles'' --- that have been shown to be
responsible for information transfer and modification~\cite{lizier2012local}.

To study emergence in GoL, a `particle collider' was considered in which two
particles are set in a colliding course~\cite{rosas2020reconciling}. The
emergent feature considered, $V_t$, was a symbolic, discrete-valued vector that
encodes the type of particle(s) present in the board. The \phiid framework (in
particular, practical criteria discussed in the previous section) provided a
quantitative validation that particles have causally emergent properties, in
line with widespread intuition, and further analyses suggested that they may be
causally decoupled with respect to their substrate.

Another demonstration of the power of the framework and practical criteria was
carried out in a computational model of flocking
birds~\cite{reynolds1987flocks,seth2010measuring}, another often-cited example
of emergent behaviour whereby the flock as a whole arises from the interactions
between individuals~\cite{rosas2020reconciling}. Here, the framework showed
that the centre of mass can predict its own dynamics better than what can be
explained from the behaviour of individual birds.

\subsection{Causal emergence in the brain}\label{sec:fmri_studies}

Moving from simulations to empirical data, the \phiid framework for causal
emergence was also adopted to study how motor behaviour might be emergent from
brain activity. Simultaneous electrocorticogram (ECoG) and motion capture
(MoCap) data of macaques performing a reaching task were analysed, focusing on
the portion of neural activity encoded in the ECoG signal that is relevant to
predict the macaque’s hand position. Results indicated that the
motion-related signal is an emergent feature of the macaque’s brain
activity~\cite{rosas2020reconciling}.

In the human brain, functional magnetic resonance imaging (fMRI) makes it
possible to study non-invasively the patterns of coordinated activity that take
place between brain regions. \phiid has been recently adopted to advance the
study of brain dynamics, moving beyond simple measures of time-series
similarity (e.g. Pearson's correlation or Shannon's mutual information) to
`information-resolved' patterns in terms of \phiid atoms. Remarkably, analyses
of human fMRI data have identified a gradient with redundancy-dominated sensory
and motor regions at one end, and synergy-dominated association cortices
dedicated to multimodal integration and high-order cognition at the other
end~\cite{luppi2020synergistic}. Recapitulating the hierarchical organisation
of the human brain, the synergy-rich regions of the human brain also coincide
with regions that have undergone the greatest amounts of evolutionary
expansion~\cite{luppi2020synergistic}. Synergy is also more prevalent as a
proportion of total information in the brains of humans than in the brains of
macaques, corroborating the link between synergistic neural information,
evolution, and higher cognitive capacities.

In this analysis the synergistic information is quantified in terms of
$\mathcal{G}^{(k)}(\bm X_t; \bm X_{t'})$ (see
Eq.~\eqref{eq:modes_of_emergence}) with $k=1$ calculated over the joint
dynamics of pairs of brain areas,\footnote{The analysis focuses on pairs of
areas because currently there is a lack of efficient estimators of
$\mathcal{G}^{(k)}(\bm X_t; \bm X_{t'})$ for three or more time series.
Developing such estimators is an important avenue for future work.} which
corresponds to the capacity of those dynamics for causal decoupling (see
Section~\ref{sec:theory}). Therefore, the results reported in
Ref.~\cite{luppi2020synergistic} indicate that causal emergence (decoupling)
increases both along the cortical hierarchy of the human brain, and across the
gap from non-human primates to humans --- with evidence suggesting that it may
support humans' ability for complex, abstract reasoning.

Relatedly, there has been a long-standing debate on whether consciousness could
be viewed as an emergent phenomenon enabled by the complex interactions between
neurons. The framework presented here provides ideal tools to rigorously and
empirically tackle this question. Moreover, causal decoupling is one of the
information atoms of a putative measure of consciousness known as
\emph{integrated information}~\cite{mediano2019beyond}, which associates the
ability to host consciousness with the extent to which a system's information
is `greater than the sum of its parts'~\cite{balduzzi2008integrated}.
Interestingly, analysis of fMRI data showed that loss of consciousness due to
brain injury corresponds to a reduction of integrated information and causal
emergence in the brain ~\cite{luppi2020synergisticb}. In this
way, the more nuanced view on neural information dynamics offered by \phiid
offers the promise of further insights for our understanding of consciousness
as an emergent phenomenon~\cite{luppi2021like}.

\section{Conclusion}

This article presents a review on how recent developments on information
decomposition naturally lead to a formal theory of causal emergence. Although
this mereological approach to causal emergence is one of many within a rapidly
growing field, it has already shown wide applicability across diverse
scientific questions. Therefore, the present review sought to bring together
the technicalities of the formalism, its interpretation, and results of its
practical application, so that each may inform the understanding of the other.

One special feature of this framework is how it allows practical criteria that
are applicable to relatively large systems, which opens a broad range of
applications. However, these tools require an explicit feature of interest,
whose definition may not be clear in some scenarios of interest (e.g. in
resting-state fMRI data). This limitation can be avoided by calculating the
capacity of emergence of the dynamics, but the calculation of this scales
poorly with the system size --- making the calculation of the emergence
capacity of large systems (such as highly multivariate brain data) currently
unfeasible. Developing procedures to either identify emergent features, or to
efficiently calculate emergent capacity in large systems are important avenues
for future work.

We hope that the theoretical and empirical advances reviewed in this article
may stimulate the growing scientific interest on emergence, which may lead the
way towards future breakthroughs on major questions about the role of emergence
in the natural world.

\vskip6pt

\section*{Acknowledgements}

We thank Joe Dewhurst, Erik Hoel, and Thomas Varley for useful discussions.

F.R. is supported by the Ad Astra Chandaria foundation. P.M. and D.B. are
funded by the Wellcome Trust (grant no. 210920/Z/18/Z). A.L. is funded by the
Gates Cambridge Trust.

\bibliographystyle{RS}

\bibliography{main}

\end{document}